\documentclass[12pt]{article}
\begin{document}
\input epsf
\def\be{\begin{equation}}
\def\bea{\begin{eqnarray}}
\def\ee{\end{equation}}
\def\eea{\end{eqnarray}}
\def\d{\partial}
\def\la{\lambda}
\def\eps{\epsilon}
\def\a{{\cal A}}
\def\tildr{\tilde}
\thispagestyle{empty}
\begin{flushright}
OHSTPY-HEP-T-02-013\\
hep-th/0211292
\end{flushright}
\vspace{20mm}
\begin{center}
{\LARGE What is the gravity dual of a chiral primary?}
\\
\vspace{20mm}
{\bf  Oleg Lunin${}^1$,  Samir D. Mathur${}^2$ and Ashish Saxena${}^2$\\}
\vspace{4mm}
${}^1${\it{School of Natural Sciences\\
Institute for Advanced Study,\\
Princeton, NJ 08540 USA}
\bigskip

${}^2$Department of Physics,\\ The Ohio State University,\\ Columbus, OH
43210, USA\\
}
\vspace{4mm}
\end{center}
\vspace{10mm}
\begin{abstract}

In the AdS/CFT correspondence a chiral primary is described  by a supergravity
solution with mass equaling angular momentum.  For $AdS_3\times S^3$
we are led to consider three
special families of metrics with this property:  metrics with conical defects,
Aichelburg-Sexl type metrics generated by rotating particles, and
metrics generated by giant gravitons.
We find that the first two of these are special cases of the complete
family of chiral primary metrics which can be
written down using the general solution in hep-th/0109154, but they
correspond to two extreme limits  -- the conical
defect metrics map to CFT states generated by twist operators that
are all identical, while the Aichelburg-Sexl metrics
yield a wide dispersion in the orders of these twists.  The giant
graviton solutions in contrast do not represent
configurations of the D1-D5 bound state; they correspond to
fragmenting this system into two or more pieces. We look
at the large distance behavior of the supergravity fields and  observe
that the excitation of these fields is  linked to the
existence of {\it dispersion} in the orders and spins of the twist
operators creating the chiral primary in the CFT.

\end{abstract}
\newpage
\setcounter{page}{1}
\section{Introduction}
\renewcommand{\theequation}{1.\arabic{equation}}
\setcounter{equation}{0}

\subsection{The issue}

Consider the AdS/CFT duality map when the spacetime is $AdS_3\times
S^3\times M_4$. The dual theory is a 1+1 dimensional CFT obtained from the
low energy limit of the D1-D5 bound state \cite{maldacena}.

More specifically, consider `global $AdS_3$' $\times S^3\times M_4$. The
state of the
CFT dual to this geometry is the Neveu-Schwarz (NS) vaccum $|0\rangle _{NS}$.
This vacuum is the simplest state of the CFT, with $h=j=\bar h=\bar j=0$. The
next simplest states are expected to be the chiral primaries of the CFT, with
$h=j, \bar h=\bar j$. What are the metrics dual to these states of the CFT?

By the general nature of the AdS/CFT map we expect that the dual of a  chiral
primaries will be created by putting an appropriate massless particle from the
supergravity multiplet into the geometry; the particle rotates on $S^3$
with angular  momentum $(j,\bar j)$ and sits at the
origin of
$AdS_3$ (so that we get `minimum energy for given charge').  In
\cite{exclusion}
the wavefunction for such a quantum was computed -- it was given by an
appropriate spherical harmonic on $S^3$ and was localized around $r=0$ in
$AdS_3$.

In the CFT, we can list chiral primaries most easily at the `orbifold
point'. At this
point in moduli space the CFT is described by a $N=4$ supersymmetric sigma
model with target space $(M_4)^N/S_N$, the symmetric product of $N=n_1n_5$
copies of $M_4$ \cite{sw}. The state dual to the
supergravity quantum mentioned above is
given by a `twist' operator acting on the  vacuum\footnote{The twist
operator $\sigma_n$ creates a cyclic permutation of   $n$ copies of
$M_4$ around
its insertion point.  The superscripts $\{--\}$ indicate that this
chiral primary
corresponds to the
$(0,0)$ form on $M_4$; action of fermion zero modes can raise the indices upto
$\{++\}$ which corresponds to the top form $(2,2)$ on $M_4$. For a
detailed study
of these chiral primaries see \cite{lm2, lm4}.}
\be
\sigma_n^{--}|0\rangle_{NS}
\ee
The operator $\sigma_n^{--}$ has $h=j=\bar h=\bar j={n-1\over 2}$.

We are interested in studying the metrics dual to the chiral primaries. Thus we
want the backreaction of the supergravity quanta on the geometry. This
backreaction will be described classically when we consider a large
number of massless quanta orbiting the $S^3$. In the CFT we expect that the
corresponding metric will be dual to a chiral primary
\be
\sigma_{n_1}^{--}\sigma_{n_2}^{--}\dots \sigma_{n_k}^{--}|0\rangle_{NS}
\label{one}
\ee

Let us see what we expect to be the nature of the corresponding geometry. A
priori, we are led to consider three possibilities:

\bigskip

(a)  {\it Conical defect metrics:}\quad We can regard the supergravity
theory as a 2+1 dimensional gravity theory arising from dimensional reduction
on
$S^3\times M_4$. A particle with angular momentum on $S^3$ appears as a
particle with mass
$m=$ in  2+1 dimensions. But in 2+1 gravity massive particles create `conical
defect metrics' so we may expect that the metrics dual to chiral primaries
(\ref{one}) will have such a conical defect structure in some way.

\bigskip

(b) {\it Aichelburg-Sexl metric:}\quad  Now consider the particles from the
viewpoint of 5+1 dimensional
$AdS_3\times S^3$ spacetime. A  chiral primary state  has the
least energy for given angular momentum. At least naively, this suggests that
we must have all the massless particles rotating on a common diameter of the
$S^3$, at the origin in $AdS_3$.  If we go close to this diameter, we
just see a uniform line of massless
particles, moving at the speed of light along this line, in a spacetime that is
essentially $5+1$ dimensional flat spacetime in the neighborhood of the line.
The metric of a massless particle in flat space is the Aichelburg-Sexl metric,
which is readily extended to a line of massless particles.  We thus
expect the metric to be
this Aichelburg-Sexl type  near the diameter,  while going over to
$AdS_3\times S^3$ at infinity.

But such a metric does {\it not} have a conical defect structure; metrics in
dimensions higher than 2+1 have power law behavior near the singularity.
Thus if such a solution were the correct one then it would certainly
be different
from (a) above.

\bigskip

(c) {\it Giant gravitons:}\quad There is yet a third possibility for
the metric that
we need to consider. It was shown in \cite{giant} that massless quanta rotating
on the sphere with high angular momenta can expand to large radius objects
while maintaining the condition $\Delta=J$. In other examples of $AdS_n\times
S^m$ spacetimes the radius of the `giant graviton' increases with its angular
momentum $J$, but for
$AdS_3\times S^3$ we have a slightly different situation.  Giant gravitons can
exist only at certain points in the moduli space. Such a special
point is obtained
for example in the case when $M_4=T^4$ when all the gauge fields are
set to zero.
At these special points a graviton can become `giant' whenever its angular
momentum $J$ is a multiple of $n_5$.  At such $J$ the potential for
the radial size
of the graviton is `flat' so the graviton can have any diameter while
maintaining
$\Delta=J$.

Geometries with a classical value for total angular momentum will
have $J\gg n_5$,
so we can obtain the desired angular momentum by using  a large number of
giant gravitons. But since these objects are extended, we expect to
get a  metric
that differs from both the metrics suggested in (a) and (b) above.

\bigskip

We can thus restate our basic question as: Which of these three kinds
of metrics
correspond to chiral primaries of the CFT?

It is possible to construct explicitly metrics of {\it each} of the
above three types.
Each metric is a classical solution of the full supergravity equations, and has
energy and angular momentum which translate to $\Delta=J$. This sharpens the
above question -- what is the correct dual of the chiral primaries of the CFT?

\subsection{Results}

In short, what we find is the following. We can construct
metrics dual
to {\it all} chiral primaries of the CFT using the solutions for the
R sector found in \cite{lm4}. From this set we see that
the `conical defect
metrics' (case (a)) are a special case\footnote{This subclass of
metrics was discussed (from the viewpoint of the
Ramond (R) sector) in \cite{bal, mm}.} where all the $n_i$ in (\ref{one})
are {\it equal}.    We then find that the Aichelburg-Sexl
type metrics (case (b)) arise from chiral primaries where a fraction
of the $n_i$
in (\ref{one}) are unity ($\sigma_1^{--}$ is just the identity
operator) while the
rest are large numbers but not necessarily equal.

Both the above metrics describe chiral primaries of the D1-D5 bound state; the
two cases give in some sense two opposite limits of the general metrics dual to
chiral primaries. By contrast, the giant graviton metrics turn out to describe
D1-D5 systems that are `disassociated' -- some fraction of the D1 and D5 branes
are pulled out from the bound state to a finite distance away from the other
branes.

In more detail, our methods and results are as follows:

\bigskip

(a)\quad We first recall the construction of \cite{lm4} that gives
the metrics for general R sector ground states. We
also recall the notion of \cite{bal,mm} that spectral flow to the NS
sector is accomplished by a `large change' of coordinates
which returns us to the same metric but with a different periodicity
for the fermions. Thus we obtain the metrics for
general chiral primary states;  for special families of states  this
construction was considered in \cite{lm5}.

\bigskip

(b)\quad We then consider two special families of metrics: the
conical defect metrics considered in \cite{bal,mm} and the
Aichelburg-Sexl type metric constructed in \cite{lm5}.  The chiral
primary state corresponding to the conical defect
metrics was constructed in \cite{lm4}, and was given by the action of
twist operators on the NS vacuum with all twist
operators having the same order and spin.  The CFT description for
the Aichelburg-Sexl metric was not known, and we
find this description here; we map the D1-D5 state by S,T dualities
to a fundamental string carrying vibrations and then,
following the general analysis of \cite{lm4}, read off the orders and
spins of the twist operators in the CFT state from the
Fourier transform of the vibration profile.

\bigskip

(c)\quad We note that we can separate a D1 brane from the D1-D5 bound
state at special values of the moduli, and
displace it away from the rest of the bound state  in a transverse
direction. This is a static configuration,
but after the coordinate change to the NS sector the separated D1
brane is seen to spin around the $S^3$ giving a giant
graviton. We get a unified picture for giant gravitons on the $AdS_3$
and on the $S^3$, as both arise as special limits of a
giant graviton that winds on both the $AdS_3$ and the $S^3$.

Using this picture for giant gravitons we can construct the metric for
any distribution of giant gravitons, since the classical metric
produced by  a set of disjoint D1-D5 systems can be written down by
just superposing the harmonic functions appearing
in the metric.  As an example we take a particular distribution of
giant gravitons and compute the metric it produces.

Given this construction of giant gravitons, we suggest that giant
graviton excitations do not describe chiral primary
states of the D1-D5 system, since such chiral primaries should be
given by configurations of the D1-D5 {\it bound} state;
the metrics for all such bound states can be obtained as in  (a) above.

\bigskip

(d)\quad We return to the metrics for generic chiral primary states
and analyze the large $r$ behavior of the first few
corrections to the background $AdS_3\times S^3$.  We note that the
conical defect metrics are locally exactly
$AdS_3\times S^3$, so all excitations save some global pure gauge
fields are zero at infinity.  For the Aichelburg-Sexl
type metric we find  nontrivial corrections at infinity, including
the excitation of some low mass scalars and gauge fields which we
study explicitly. We conjecture that the
existence of these fields  is related to the fact that the twist
operators creating the dual CFT state have a wide dispersion in
their orders in the Aichelburg-Sexl case, while they have no
dispersion in the conical defect metrics. Following up on this
conjecture we look at the asymptotic fields for generic chiral
primary states, and find that we can relate the existence of
these corrections to the presence of dispersions in the orders and
spins of the twist operators creating the dual CFT state.

\bigskip

\section{The metrics for general chiral primaries}
\renewcommand{\theequation}{2.\arabic{equation}}
\setcounter{equation}{0}

\subsection{ The D1-D5 and FP systems}

Let us recall some results from \cite{lm3, lm4, lm5}. We take $M_4=T^4$ for
concreteness.  Consider type IIB string theory compactified on
$T^4\times S^1$.  The  $T_4$ has volume $(2\pi)^4 V$ and  is parametrized by
coordinates
$z_1, z_2, z_3, z_4$. The $S^1$ has radius
$2\pi R$ and is parametrized by a coordinate $y$. The noncompact directions are
$x_0, x_1, x_2, x_3, x_4$.  We wrap $n_5$ D5 branes on $T^4\times S^1$ and
$n_1$ D1 branes on $S^1$. We are interested in the CFT arising from the low
energy physics of the {\it bound state} of these branes.

The field theory describing this bound state has its fermions
periodic around the
$S^1$ and is thus in the Ramond (R) sector.\footnote{The periodicity of the
worldvolume fermions in induced from the periodicity of fermions in the bulk,
and in the bulk  the fermions must be periodic in order that the cosmological
constant vanish and flat spacetime be a solution at infinity.} The R ground
state of the CFT is highly degenerate, with a degeneracy $\sim
e^{2\sqrt{2\pi}\sqrt{n_1n_5}}$.

One might naively think that such bound states are
pointlike in the transverse space $x_1, x_2, x_3, x_4$, and that the
corresponding metrics have a pointlike singularity at
the origin $r=0$ in this space. To see that such is not the case and
to construct the actual
metrics for the
D1-D5 ground states, it is helpful to map the D1-D5 system by a set of $S,T$
dualities to the FP system:  we obtain a fundamental string (F) wrapped $n_5$
times around $S^1$, carrying momentum (P) along the $S^1$. The fact that we
are studying a {\it bound} state of these charges implies that the F
string is a
single `multiwound' string, and all the momentum P is carried by traveling
waves on this string.

It is possible to write down the supergravity solution for a string
carrying traveling waves in one direction. The only subtlety is that we have
many strands of the `multiwound' string, and must superpose harmonic functions
arising from different strands.\footnote{The different stands do not
all carry the
same profile of vibration, since the only restriction on the
vibration is that it
close after $n_5$ turns around the $S^1$.  But as long as the momentum flows
along the same direction on each strand we can superpose harmonic
functions, and   the final configuration is a  1/4 BPS
solution of IIB supergravity.} Such a superposition was carried out in
\cite{lm3} for the case where the string profile was a uniform helix; dualizing
back to the D1-D5 system we obtained the subclass of metrics studied
in \cite{bal, mm, cveticyoum}. More generally, let the F string  have a
vibration profile in the
noncompact spatial directions\footnote{We do not consider vibrations in the
$T^4$ directions, though in principle these could be handled the same way. The
restriction on vibrations corresponds, after mapping to the D1-D5 system, to
using cohomology elements $h_{00}, h_{02}, h_{20}, h_{22}$ from $T^4$. Both
$T^4$ and $K3$ have the one form of each of these types; the remaining
vibrations would differentiate between these two spaces.} described by
$x_i=F_i(v)$, where
$i=1,2,3,4$ and
$v=t-y$.   The supergravity solution created by such a string is
given in eqn. (\ref{ChiralNull}) in Appendix A.
If we perform the chain of dualities that take us to the D1-D5 system
then we get the supergravity solution
\bea\label{D1D5Chiral}
ds^2&=&\sqrt{\frac{H}{1+K}}\left[-(dt-A_idx^i)^2+(dy+B_idx^i)^2\right]
+
\sqrt{\frac{1+K}{H}}d{\vec x}\cdot d{\vec x}\nonumber\\
&+&\sqrt{H(1+K)}d{\vec z}\cdot d{\vec z}\\
e^{2\Phi}&=&H(1+K),\qquad
C^{(2)}_{ti}=\frac{B_i}{1+K},\qquad
C^{(2)}_{ty}=-\frac{K}{1+K},\nonumber\\
C^{(2)}_{iy}&=&-\frac{A_i}{1+K},\qquad
C^{(2)}_{ij}=C_{ij}+\frac{A_iB_j-A_jB_i}{1+K}
\eea
The functions $H$, $K$ and $A_i$ appearing in this solution are
related to the profile ${\bf F}(v)$:
\bea\label{CNMProf}
H^{-1}=1+\frac{Q}{l}\int_0^l\frac{dv}{({\vec x}-{{\vec F}})^2},\quad
K=\frac{Q}{l}\int_0^l\frac{|\dot F|^2dv}{({\vec x}-{{\vec F}})^2},\quad
A_i=-\frac{Q}{l}\int_0^l\frac{{\dot F}_i dv}{({\vec x}-{{\vec F}})^2}
\eea
and the forms $B_i$ and $C_{ij}$ are defined by the duality relations:
\be\label{DualFields}
dB=-^*dA, \qquad dC=-^*dH^{-1}.
\ee
(The form $C^{(2)}$ is the Ramond-Ramond $B$ field $B^{RR}_{ij}$.)

Different profiles of the vibrating F string correspond to different R ground
states of the D1-D5 system. (This relation was studied in detail in
\cite{lm4} and
will be reviewed below.) It is important that the F string has no longitudinal
vibration mode, and must therefore necessarily bend away from the  central
axis in order to carry the momentum P. The corresponding metrics thus have an
extended singularity rather than a pointlike singularity at $r=0$.

\subsection{ Spectral flow}

As mentioned above, the above metrics describe  states of the R sector,
since the  fermions are periodic under
$y\rightarrow y+2\pi R$. We wish to obtain metrics for the NS sector, where we
have chiral primaries ($h=j$, $\bar h=\bar j$). The operation of spectral flow
maps R sector ground states in the CFT to chiral primaries in the NS sector of
the CFT, in a 1-1 and onto fashion.  For the special class of D1-D5 metrics
studied in \cite{bal, mm} the description of spectral flow in the supergravity
dual was discussed, and we can extend that to the general set of
metrics (\ref{D1D5Chiral}) as follows.

We first take the limit
\be
R\gg (Q_1Q_5)^{1/4}\gg a, ~|x|\ll (Q_1Q_5)^{1/4},
\label{five}
\ee
where $a$ is the size of the singularity ($|{\bf F}|<a$).
Then we should put the  following expressions for
$H^{-1}$ and $1+K$ (since unity can be ignored in comparison to the
term of the form $\sim 1/x^2$)
\be\label{AdSlimit}
H^{-1}=\frac{Q}{l}\int_0^l\frac{dv}{({\vec x}-{{\vec F}})^2},\quad
1+K=\frac{Q}{l}\int_0^l\frac{|\dot F|^2dv}{({\vec x}-{{\vec F}})^2}
\ee
in (\ref{D1D5Chiral}) and we get another set of  exact solutions to
supergravity. The region
\be
a\ll x\ll (Q_1Q_5)^{1/4}
\label{eone}
\ee
is the large radius region of anti-de-Sitter space
and here the metric is locally $AdS_3\times S^3\times T^4$.  To see
this explicitly,
consider the four coordinates
$x_1, x_2, x_3, x_4$ and map these to a set of polar coordinates
\bea\label{EpolarMap}
x_1&=&{\tilde r} \sin{\tilde \theta} \cos{\tildr\phi}, ~~x_2={\tilde r}
\sin{\tilde\theta} \sin{\tildr\phi},\nonumber \\
x_3&=&{\tilde r} \cos{\tilde\theta} \cos{\tildr\psi}, ~~x_4={\tilde r}
\cos{\tilde\theta} \sin{\tildr\psi}
\label{etwo}
\eea
In the region (\ref{eone}) the 3-dimensional surface of constant
${\tilde r}$ is an $S^3$ with proper radius $(Q_1Q_5)^{1/4}$
\be
ds^2=(Q_1Q_5)^{1/4}[d{{\tilde\theta}}^2+\sin^2{\tilde\theta}
d{\tildr\phi}^2+\cos^2{\tilde\theta} d{\tildr\psi}^2]
\ee
\medskip

The fermions have charge $({1\over 2}, {1\over 2})$ under the
$SU(2)\times SU(2)\approx SO(4)$ symmetry
group of the $S^3$.  In the 1+1 dimensional CFT we can go from the R
to the NS sector if we add an extra
flat connection $A_y$ with $e^{i\int A\cdot dl }=-1$ in each $SU(2)$.  Such a
connection can be induced by a coordinate transformation which rotates the
$S^3$ as we move along the $y$ circle.  In the supergravity dual such
a rotation of the $S^3$ induces the map
\bea\label{RtoNS}
{\psi}={\tildr\psi}-\frac{y}{R}\qquad
{\phi}={\tildr\phi}-\frac{t}{R}
\label{efour}
\eea
This generates the appropriate connection $A$, and the fermions may now be
considered to be in the NS sector.\footnote{In the NS sector we will
often write $\chi=y/R$.}
    For the subclass of
metrics in
\cite{bal, mm}
the corresponding NS sector metrics were studied in \cite{lm5}. But we may
make the same coordinate change at infinity also for the more general class
(\ref{D1D5Chiral}), and thus obtain metrics dual to all chiral primary states
of the NS sector.

\subsection{The detailed FP $\rightarrow$ D1-D5 map}

We recall in  more detail the map \cite{lm4} between 1/4 BPS
states of the FP system and chiral primaries   of the
D1-D5 system.  The traveling waves travel in only one direction
along the F string, for states with supersymmetry. The
string has a total length $2\pi R' n_5$, where $R'$ is the radius of
the $S^1$. On this string the traveling wave can be
described by specifying the number of quanta in each harmonic.
Creating a quantum in the $n$th harmonic implies in the
D1-D5 system
            the action of the twist operator $\sigma_n$ on $|0\rangle_{NS}$:
\be\label{theMap}
a^{\dagger,i}_n~\rightarrow ~\sigma_n^{\epsilon, \epsilon'}
\label{three}
\ee
Here $i=1,2,3,4$ labels the four possible polarizations of the
vibration mode in the four noncompact directions $x_1, x_2,
x_3, x_4$.   We can write
$SO(4)\approx SU(2)\times SU(2)$ and thus re--express this vector index
$i$ in terms of a pair of spinor indices
$\epsilon=\pm,
\epsilon'=\pm$ which describe the spin $({1\over 2}, {1\over 2})$
representation  of $SU(2)\times SU(2)$. In the D1-D5
system in the NS sector we have four chiral primaries for each twist
$\sigma_n$ :
\bea
\sigma_n^{--},~~h=j_3={n-1\over 2}, ~\bar h=\bar j_3={n-1\over 2}\\
\sigma_n^{+-},~~h=j_3={n+1\over 2}, ~\bar h=\bar j_3={n-1\over 2}\\
\sigma_n^{-+},~~h=j_3={n-1\over 2}, ~\bar h=\bar j_3={n+1\over 2}\\
\sigma_n^{++},~~h=j_3={n+1\over 2}, ~\bar h=\bar j_3={n+1\over 2}
\label{four}
\eea
These  twist operators are thus also labelled by a set
$\epsilon=\pm, \epsilon'=\pm$,  The map (\ref{three}) says
that we should identify the indices $(\pm, \pm)$ on the $\sigma_n$
with the indices
$(\pm, \pm)$ obtained from the vector index $i$ on the
$a^{\dagger,i}_n$. The reason for this is the following.
The S,T dualities relate the F string carrying the vibrations
$a^{\dagger,i}_n$ to the D1-D5 system
in the R sector. If we spectral flow the chiral primaries
$\sigma_n^{\epsilon, \epsilon'}$ from the NS to the R sector then
we find that they have
      $SU(2)\times SU(2)$ spins
\be
j_3={\epsilon\over 2}, ~~\bar j_3={\epsilon'\over 2}
\ee
and they form the $({1\over 2}, {1\over 2}) $ representation of
$SU(2)\times SU(2)$.  The spins of the vibrations on the F string and the spins
in the R sector of the D1-D5 system are immediately identified with
each other, since the S,T dualities
do not affect the transverse coordinates $x_1, x_2, x_3, x_4$ which
give the $SO(4)\approx SU(2)\times SU(2)$.

We can  get a classical vibration profile for the string if we excite
a large number of vibration quanta.  Such a state maps to a
state of the form (\ref{one}) in the NS sector of the D1-D5 system.
\bigskip

\section{Metrics for special cases of interest}
\renewcommand{\theequation}{3.\arabic{equation}}
\setcounter{equation}{0}

The solutions (\ref{D1D5Chiral}) possess a singularity at the location of the F
string in the FP system, which maps to a corresponding
singularity in the D1-D5 system. It is important to note that this
singularity is  an acceptable one that will be resolved in the
full string theory.  In the FP system we have started with a
fundamental string in flat space, and this string is a valid
quantum in the full string theory. Thus the metric it creates is a
true solution of the full theory. The dualities we have used
to go to the D1-D5 system are also true dualities since they are made
along closed cycles of the $T^4$; thus they are not
just `solution generating techniques' for  classical supergravity
solutions, which would be the case for example if we
dualized along some other direction of the spacetime. In performing
this duality we have `smoothed over' the singular locus
of the F string since this string had a large number of strands (this
is explained in detail in \cite{lm4});  this smoothing
should generate the correct classical D1-D5 solution for the R
sector. The truncation to the $AdS$ region was done by a
limiting process $R\gg (Q_1Q_5)^{1/4}$ and looking only at $r\ll
(Q_1Q_5)^{1/4}$; this
is therefore also an allowed step in the full quantum
string theory. Lastly,  map to the NS sector was just  a change of
coordinates (\ref{RtoNS}), and so does not spoil the existence of
the solution. Thus the metrics we make for the chiral primaries are
the classical limits of exact string theory solutions, and
not just classical solutions with singularity.

\subsection{Case I:  $AdS_3\times S^3$ }

Let the F string carry a traveling wave which has all its energy
only in the lowest harmonic. Further, let the polarization of
this wave be such that the string rotates in a helical fashion in the
$x_1-x_2$ plane (recall that $x_1, x_2, x_3, x_4$ are the
four transverse direction in which we are considering vibrations). It
was shown in \cite{lm3} that this configuration of the
FP system possesses the maximal possible angular momentum for given F
and P charges: under the $SU(2)\times SU(2)$
this configuration has $(j,j')=({n_1n_5\over 2}, {n_1n_5\over 2})$.
The metric of this configuration was computed, and was
shown to map under dualities to the `maximally rotating' D1-D5 bound
state discussed in \cite{bal, mm}.

In more detail, consider a  FP metric generated by a following profile:
\be
G_1=a\cos\omega v',\quad G_2=a\sin\omega v', \quad G_3=G_4=0
\label{yyonePrime}
\ee
We choose
\be
\omega=\frac{1}{n_5R'}
\ee
which corresponds to the F string  having all its energy in the lowest
harmonic (recall that the F string has winding number $n_5$ around
the $y$ circle, and we use $R'$ to denote the radius of $y$ direction on
the FP side).

We
wish to write the metric for the corresponding D1-D5 system. In terms of the
parameters of the D1--D5 system the profile (\ref{yyonePrime}) becomes:
\be
F_1=a\cos\omega v,\quad F_2=a\sin\omega v, \quad F_3=F_4=0,\quad
\omega=\frac{R}{n_5}
\label{yyone}
\ee
For later use it is convenient to introduce
\be
l=2\pi R' n_5=\frac{2\pi n_5}{R}
\ee
The D1-D5 metric is given by eqn. (\ref{D1D5Chiral}), with
coefficient functions
$H^{-1}, K, A_i$ given by (\ref{CNMProf}).   As an example of the
computation note that in $H^{-1}$ we get
     the harmonic function created by a uniform circular source:
\bea
H^{-1}&=&1+{Q\over 2\pi}\int_0^{2\pi} {d\xi\over
(x_1-a\cos\xi)^2+(x_2-a\sin\xi)^2+x_3^2+x_4^2}\nonumber\\
&=&1+{Q\over
\sqrt{(\tilde r^2+a^2)^2-4 a^2\tilde r^2\sin^2\tilde\theta}}
\eea
where in the last step we have used the polar coordinates
(\ref{etwo}). The above expression simplifies if we
change from $\tilde r, \tilde\theta$ to coordinates $r,\theta$:
\bea
{\tilde r}&=& \sqrt{r^2+a^2\sin^2\theta}, ~~\cos{\tilde\theta}
={r\cos\theta\over \sqrt{r^2+a^2\sin^2\theta}}
\label{ethree}
\eea
(${\tildr\phi}$ and ${\tildr\psi}$ remain unchanged). Then we get
\be
H^{-1}=1+{Q\over r^2+a^2\cos^2\theta}
\ee

We can go to the near horizon limit ($r\ll
(Q_1Q_5)^{1/4}$),  and then the metric of
the D1--D5 system becomes\footnote{The complete asymptotically flat
metric is given for example in
\cite{lm3}.} (we write $r'=r/a$):
\bea
ds^2&=&-({r'}^2+1)\frac{a^2dt^2}{\sqrt{Q_1Q_5}}+{r'}^2
\frac{a^2dy^2}{\sqrt{Q_1Q_5}}+
\sqrt{Q_1Q_5}\frac{d{r'}^2}{{r'}^2+1}\nonumber\\
&+&\sqrt{Q_1Q_5}\left[d\theta^2+\cos^2\theta \left(d{\tildr\psi}-
\frac{ady}{\sqrt{Q_1Q_5}}\right)^2+
\sin^2\theta \left(d{\tildr\phi}-\frac{adt}{\sqrt{Q_1Q_5}}\right)^2\right]
\eea
    The charge $Q_1$ is related by dualities to the momentum charge P
carried by the F string,
and for the profile (\ref{yyone}) we find
\be
Q_1=a^2\omega^2Q_5,
\ee
which for the profile we are considering translates into
\be
\frac{a}{\sqrt{Q_1Q_5}}=\frac{1}{R}.
\ee
Lastly, performing the coordinate change (\ref{efour}) that gives spectral flow
we get the metric
\bea
\label{esix}
ds^2&=&\sqrt{Q_1Q_5}\left[
-({r'}^2+1)\frac{dt^2}{R^2}+{r'}^2
\frac{dy^2}{R^2}+
\frac{d{r'}^2}{{r'}^2+1}\right]\nonumber\\
&+&\sqrt{Q_1Q_5}\left[d\theta^2+\cos^2\theta d{\psi}^2+
\sin^2\theta d{\phi}^2\right],
\eea
which is just  $AdS_3\times S^3\times T^4$.

On the CFT side, the fact that all the energy of the FP system is in
the lowest harmonic implies under the map (\ref{theMap}) that all
the twist operators have order unity. The choice of rotation plane
($x_1-x_2$) implies that we get the operator
$\sigma_1^{--}$ for each quantum of vibration. But this is just the
identity operator, with $h=j=\bar h=\bar j=0$. Thus the
CFT state dual to the metric under consideration is (in the NS
sector) just $|0\rangle_{NS}$.

In Figure 1(a) we have sketched the F string, opened up to its full
length $2\pi R' n_5$ and carrying the vibration profile
mentioned above. The dual state $|0\rangle_{NS}$ is of course the
simplest chiral primary. We will use similar figures to discuss
more complicated cases below.

\begin{figure}\hspace{1in}
\epsfysize=3in \epsffile{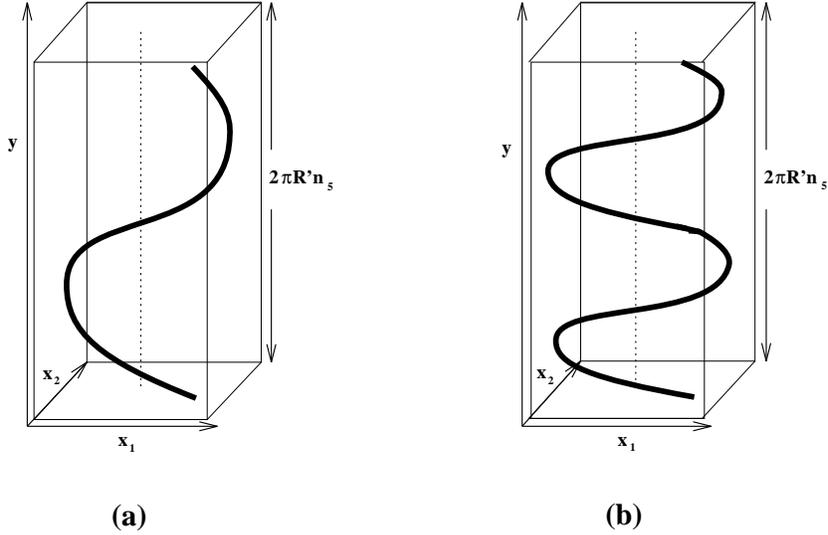}
\caption{\label{fig1}
(a) The F string opened up to its full length, rotating in a
helix with a single turn; this gives $AdS_3\times S^3$.
(b) The F string executing a helix in the second harmonic; this
gives a conical defect metric.
}
\end{figure}

\subsection{Case II: The conical defect metrics}

Take the F string to again rotate in the $x_1-x_2$ plane but let all
the energy be in the $m$th harmonic (instead of the first
harmonic). Figure 1(b) shows the F string carrying such a vibration
with $m=2$.  The metric for this configuration yields
\cite{lm3}, after dualization to the D1-D5 system, the general
metrics discussed in
\cite{bal, mm}.  After the limit (\ref{five}) is taken,  we get a metric
that is  locally $AdS_3\times S^3\times
T^4$.  But there is a singular circle, such that at any point along
this circle we have a `conical defect': a wedge is removed
from  the $AdS_3$ and the $S^3$ is glued across the cut after a
suitable rotation. Taking the limit (\ref{five}), and making the coordinate
changes (\ref{etwo}), (\ref{ethree}), (\ref{efour})  just as in Case I
above we get the metrics for the chiral primary \cite{lm5}:
\bea
\label{two}
ds^2&=&\sqrt{Q_1Q_5}\left[
-({r'}^2+\gamma^2)\frac{dt^2}{R^2}+{r'}^2
\frac{dy^2}{R^2}+
\frac{d{r'}^2}{{r'}^2+\gamma^2}\right]\\
&+&\sqrt{Q_1Q_5}\left[d\theta^2+\cos^2\theta
\left(d{\psi}+(1-\gamma)\frac{dy}{R}\right)^2+
\sin^2\theta \left(d{\phi}+(1-\gamma)\frac{dt}{R}\right)^2\right],\nonumber
\eea
where $\gamma=1/m$.

In the CFT we find that the dual state is (using the map (\ref{theMap}))
\be\label{ChirPrCon}
[\sigma_m^{--}]^{N/m}|0\rangle_{NS},\
\ee
where $N=n_1n_5$.  The number $N/m$ of twist
operators comes from
the fact that all the $N$ copies of $T^4$ in the
symmetric product must be involved in the permutation (a copy not
involved in the permutation gives a permutation
cycle of length unity, which gives  the twist $\sigma_1$ and implies
a vibration quantum in the {\it first} harmonic on the
F string).

\subsection{Case III - the Aichelburg-Sexl metric}
\label{Sect33}

Consider the D1-D5 geometry  $AdS_3\times S^3\times T^4$; this is the
dual of the NS vacuum $|0\rangle_{NS}$.
As discussed in section 1, we can  get  a chiral
primary with $h=j=\bar h=\bar j$ by taking  massless
quanta at the center of $AdS_3$, rotating on the $S^3$. The angular
momentum on the $S^3$ is described by the spin
state under the group $SO(4)\approx SU(2)\times SU(2)$, and we should
take $|j,j_3\rangle=| j,j\rangle$, $|\bar j, \bar
j_3\rangle=|j,j\rangle$ to get the chiral primary. This implies that
the quantum rotates along the direction $\phi$ at
$\theta=\pi/2$.

We are interested in the metric obtained after we take the
backreaction of the rotating quanta into account. To reach the
classical limit we take several such massless quanta rotating along
the $\phi$ circle at $\theta=\pi/2$. These quanta can
have different values for $j$, but the overall distribution is
selected to satisfy the following conditions:

\bigskip

(i)\quad We take a large number of quanta so that we can approximate
the energy distribution by a continuous one along
the circle $\theta=\pi/2$.

\bigskip

(ii)\quad We let the distribution of quanta be uniform along the
$\phi$ coordinate, so that the density of mass and energy is
uniform along the circle $\theta=\pi/2$.

\bigskip

(iii) \quad We let each $j\gg 1$, so that the quanta have wavelength
$\lambda\ll R_{AdS}$, where $R_{AdS}=(Q_1Q_5)^{1/4}$ is
the radius of the $AdS_3$ and the $S^3$. This allows us to treat the
quanta as pointlike.

\bigskip

Near this line of massless quanta $\theta=\pi/2, r=0$ we should find
the metric produced by a uniform line of massless
particles in 5+1 flat spacetime (we always smear all wavefunctions
uniformly over the $T^4$). The latter distribution
generates the metric
\be\label{OrigAS}
ds^2=-dt^2+dz^2+\frac{q}{(x_ix_i)}(dt-dz)^2+\sum_{i=1}^4 dx_idx_i
\ee
We thus need a solution to the full supergravity equations that
behaves as (\ref{OrigAS}) near the singular line and goes over to
$AdS_3\times S^3\times T^4$ at large $r$.  There is no obvious way to
find such a solution. The general solutions (\ref{D1D5Chiral})
constructed above started with a given profile for the F string, while
in the present case we want a D1-D5 metric but do not
know which traveling wave on the F string will generate the answer.
But the Aichelburg-Sexl solution  in asymptotically flat space has an
interesting property: the linearized gravity solution turns out to
also be an exact solution. In the present case one can find
the linearized solution with some effort, and again this linearized
solution turns out to be exact. This procedure was carried
out in \cite{lm5} and the following solution was obtained for the
line of massless particles\footnote{We recall, that since we are in
the NS sector, the coordinates on the sphere are
called $\theta,{\phi},{\psi}$. In (\ref{NewASmetr}) we
also use coordinates $\tau$ and $\chi$ which are related with
$t$ and $y$: $t=\tau R$, $y=\chi R$.}:
\bea\label{NewASmetr}
ds^2&=&L^2\left\{-(1+{{r'}^2})d\tau^2+\frac{d{r'}^2}{1+{r'}^2}+
{r'}^2 d\chi^2+d\theta^2+\cos^2\theta d{\psi}^2+\sin^2\theta
d{\phi}^2\right\}
+\sum_{i=1}^4 dz_i dz_i\nonumber\\
&+&\frac{qL^2}{{r'}^2+\cos^2\theta}\left[\left\{(1+{{r'}^2})d\tau-
\sin^2\theta d{\phi}\right\}^2-
\left\{{{r'}^2}d\chi-\cos^2\theta d{\psi}\right\}^2\right],
\eea
the R--R two--form field and dilaton are:
\bea\label{NewASfield}
B^{RR}&=&L^2e^{-\Phi_0}\cos^2\theta d{\phi}\wedge
d{\psi}+{{r'}^2}L^2e^{-\Phi_0} d\tau \wedge d\chi\nonumber\\
&-&\frac{qL^2e^{-\Phi_0}}{{r'}^2+\cos^2\theta}
\left\{(1+{{r'}^2})dt-
\sin^2\theta d{\phi}\right\}\wedge\left\{
{r'}^2d\chi-\cos^2\theta d{\psi}\right\}\\
\label{NewASdilat}
e^{2\Phi}&=&e^{2\Phi_0}.
\eea
Here $e^{2\Phi_0}$ is an arbitrary constant which we will later
interpret as the ratio of
the brane charges: $e^{2\Phi_0}=Q_1/Q_5$.

Looking at this D1-D5 geometry we cannot directly see what state in
the CFT it is dual to. What we should do now is to find an FP
solution that is the dual of this D1-D5 solution; the Fourier
analysis of the FP solution using the relation (\ref{theMap}) will
then tell  us
which chiral primary the solution corresponds to.

To find the profile ${\bf F}(v)$ one should perform several steps.
First one should go from the solution in the
NS sector (\ref{NewASmetr})--(\ref{NewASdilat}) to the solution in
the Ramond sector by making a change of coordinates
inverse to (\ref{RtoNS}). From the resulting solution (which is
asymptotically $AdS_3\times S^3$) one can read off the
values
of the harmonic functions, and the extension to the asymptotically
flat space is achieved by adding a constant to $H^{-1}$.
After performing these steps we get:
\bea\label{ASCNM}
H^{-1}=1+\frac{L^2e^{-\Phi_0}}{{r}^2+a^2\cos^2\theta},\quad
K=\frac{L^2e^{\Phi_0}}{{r}^2+a^2\cos^2\theta},\quad
A_\phi=\sqrt{1-q}\frac{L^2 a\sin^2\theta}{{r}^2+a^2\cos^2\theta}
\label{jjtwo}
\eea
Here we introduced a new coordinate
\be
{r}=\sqrt{1-q}~Lr'
\ee
which becomes a radial coordinate at the flat infinity, and the parameter
\be
a=\sqrt{1-q}~L
\ee
Let us now discuss how we can get the harmonic functions
(\ref{ASCNM}) from the profile of a vibrating string.

As a warm up example, first consider  the vibration profile of the F
string shown in Fig. 2(a).  We have seen that the
total length of this string is
$2\pi R' n_5$. A part of this string (characterized by a fraction
$\xi<1$) executes a single turn around a uniform helix in the
$x_1-x_2$ plane. The remainder of the string is at a constant
location in the $x_1,x_2, x_3, x_4$ space.  Thus the profile
$F_i(v)$ is
\bea
&&{{\bf F}}(v)=a{\bf e}_1\cos\left(\frac{2\pi mv}{\xi l}\right)+
a{\bf e}_2\sin\left(\frac{2\pi mv}{\xi l}\right),\quad 0\le v<l\xi\\
&&{{\bf F}}(v)=a{\bf e}_1,\quad l\xi\le v<l,\qquad l=\frac{2\pi n_5}{R}
\label{jjone}
\eea
This profile gives rise to the coefficient functions
\bea\label{CNMInt}
H^{-1}=1+\frac{Q\xi}{{r}^2+a^2\cos^2\theta}+\frac{Q(1-\xi)}{(x_1-a)^2+
x_2^2+x_3^2+x_4^2},\\
A_{\phi}=\frac{a^2Q}{{r}^2+a^2\cos^2\theta}\frac{2\pi
}{l}\sin^2\theta,\quad
K=\frac{Qa^2}{{r}^2+a^2\cos^2\theta}\frac{1}{\xi}\left(\frac{2\pi }{l}\right)^2
\eea
\begin{figure}\hspace{1in}
\epsfysize=3in \epsffile{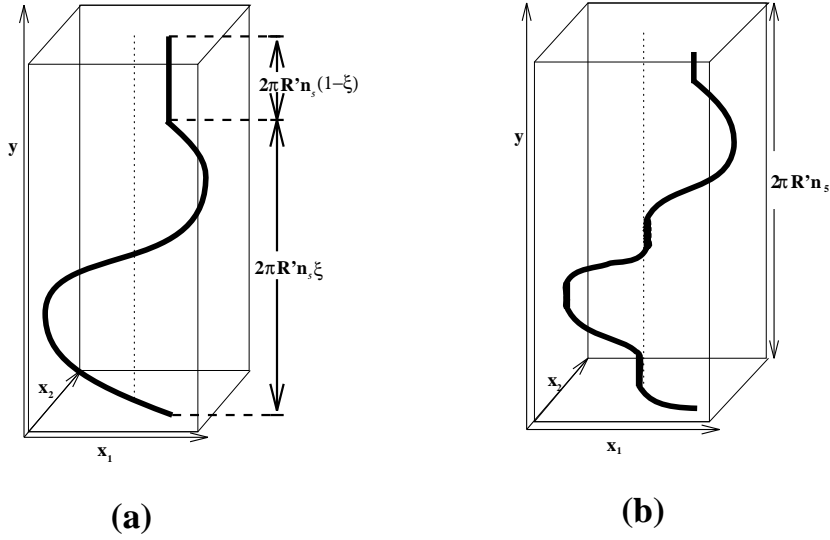}
\caption{\label{fig2}
(a) The F string executes one turn of a helix and then stays
at a constant location.  (b) The  constant part
of the string is broken into several segments which are interspersed
with the curved part; this gives the Aichelburg-Sexl
solution.
}
\end{figure}

Now consider the profile of the F string depicted in Fig 2(b).  The
helical part and the straight part in Fig 1(a) have each
been broken up into several segments and such segments alternate:
thus the string describes a part of a helix, then stays at
a constant $x_i$ location for some length, then again describes a
part of the helix, and so on.
The locations and lengths of the
straight line segments are arbitrary, except that we impose the
following overall conditions:

\bigskip

(i')\quad We take a large number of straight line segments so that we
can approximate their  distribution by a
continuous one along the circle $x_1^2+x_2^2=constant$.

\bigskip

(ii')\quad We let the distribution of straight line segments be
uniform (in a coarse grained sense) along this circle.

\bigskip

(iii') \quad We let each straight line segment have a length $l\gg 2\pi
R'$ so that it describes several windings around the
compactified circle $y$.

\bigskip

For such an F string we get a change in the last term in $H^{-1}$ in
(\ref{CNMInt}).  The location
$(a,0,0,0)$ in (\ref{jjone}) for the straight part of the F string is
now smeared uniformly over a circle of radius $a$ in the $x_1-x_2$ plane.
In the harmonic function $H^{-1}$ we get a linear superposition of
the effects of different parts of the F string,  and we find that
with this smearing we
need to   take the average of the last term in $H^{-1}$ over the
locations of the straight line segments. We get
\be
H^{-1}=1+\frac{Q}{{r}^2+a^2\cos^2\theta}
\ee
and similarly all the other  coefficient functions (\ref{jjtwo}) with following
identification of the parameters:
\be
L=(Q_1Q_5)^{1/4},\quad e^{2\Phi_0}=\frac{Q_1}{Q_5},\quad Q_5=Q,\quad
Q_1=\frac{Qa^2}{\xi}\left(\frac{2\pi}{l}\right)^2
\ee

Having obtained the  profile of the F string  which gives the
Aichelburg-Sexl solution for the D1-D5 system
we can find the corresponding
chiral primary in the CFT by the map (\ref{theMap}).
The conditions (i'), (ii'), (iii') correspond respectively to the
requirements (i), (ii), (iii) above. In particular (iii') implies that
the straight line segments give rise to chiral primaries which have
twist operators $\sigma_n$ with $n\gg 1$. The fact that the
curved part of the string describes only one cycle of the helix
implies that it contributes only copies of $\sigma_1^{--}$
which is just the identity operator and thus does not affect the NS
vacuum. Thus the state of the dual CFT has the form
\be
[\sigma_1^{--}]^p\sigma_{n_1}^{--}\sigma_{n_2}^{--}\dots
\sigma_{n_k}^{--}|0\rangle_{NS}~=~\sigma_{n_1}^{--}\sigma_{n_2}^{--}\dots
\sigma_{n_k}^{--}|0\rangle_{NS}
\label{oneState}
\ee
with all $n_i\gg 1$. Note that $\sum_{i} n_i+p=N$.

We thus see that the Aichelburg-Sexl metric, which has no conical
defect, also describes chiral primaries, but these chiral
primaries are rather different from those of the type (\ref{ChirPrCon}). In
(\ref{ChirPrCon}) all order $n_i$ of the twists are equal, so the
dispersion $\langle n_i^2\rangle-\langle n_i \rangle^2=0$. By
contrast, in (\ref{oneState}) we have $\langle n_i^2\rangle-\langle n_i
\rangle^2$ of order $\langle n_i\rangle ^2$, where we have taken into
account the fact that
a significant number $p$ of the indices involved in the permutations
are in permutation cycles of length $n_i=1$ -- i.e.,
they are left untouched by the permutation.

\section{Giant gravitons}
\renewcommand{\theequation}{4.\arabic{equation}}
\setcounter{equation}{0}

We have seen above that both the conical defect metrics of \cite{bal,
mm} and the Aichelburg-Sexl type metric found in
\cite{lm5} are special cases of the metrics that were obtained for
generic chiral primaries in \cite{lm4}. We now proceed to
examine giant gravitons and the metrics they would create by their
backreaction on the geometry.

Giant gravitons exist in $AdS_3\times S^3\times T^4$ only for a
subset of the possible values of the moduli; we can get giant
gravitons for example if we set all gauge potentials like the NS $B$
field to be zero on $T^4$. Let us assume that we have
chosen such a spacetime.  As before we smear all fields on the $T^4$,
and consider the 5+1 spacetime $AdS_3\times S^3$. We
can get giant gravitons where the D1 brane expands in this D1-D5
geometry, and also when the D5 brane expands -- the
D5 brane wrapped on $T^4$ is also a string in the 5+1 spacetime. (We
can also get all possible bound states of these two
kinds of giant gravitons, as we will see below.)

For $AdS_3\times S^3$ the  metric and RR $B$ field are\be
ds^2=-\cosh^2\rho d\tau^2+d\rho^2+\sinh^2\rho d\chi^2
+d\theta^2+\cos^2\theta d\psi^2+
\sin^2\theta d\phi^2
\ee
\be
B^{RR}=\cos^2\theta d\phi\wedge d\psi+\sinh^2\rho d\tau\wedge d\chi
\ee

Consider a `giant graviton' made of the threshold bound state of
$m_1$ D1 branes and $m_5$ D5 branes. The DBI + Chern
Simons action is (recall that the $H^{RR}$ field strength is
self-dual in 5+1 dimensions)
\bea
S&=&-(m_1T_1+m_5T_5V)\int d^2\sigma\sqrt{-G_{MN}\d X^M\d X^N}\nonumber\\
&+&
\frac{m_1 T_1+m_5 T_5V}{2}\int d^2\sigma\eps^{ab}B_{MN}^{RR}\d_a X^M\d_b X^N
\eea

We let the giant graviton run around the sphere along the coordinate
$\phi$. In other spacetimes of the form
$AdS_m\times S^n$ we have two kinds of giant gravitons: those which
expand  on the $S^n$, and those which expand on
the $AdS_m$. In the present case we see however that we can have a
more general giant graviton, which has a nonzero
size in {\it both} $AdS_3\times S^3$. Letting $\sigma^0, \sigma^1$ be
the world sheet coordinates of the giant graviton
we write the ansatz
\be
\tau=\sigma^0,\quad \chi=\sigma^1,\quad \psi=\pm \sigma^1, \quad
\rho=\bar \rho,\qquad \theta=\bar \theta,\qquad \phi=\phi(\tau).
\label{eseven}
\ee
This giant graviton winds  once around the $\chi$ circle in $AdS_3$
and also once around the $\psi$ circle in $S^3$, rotates  along
the $\phi$ direction, and maintains a constant radius $\bar\rho$ on $AdS_3$
and a constant radius $\bar\theta$ on $S^3$.
Extremsing the action yields a solution with
\be
\phi=\pm \tau=\pm t
\label{eeight}
\ee
so that the giant graviton rotates with unit angular velocity.  The
choice of sign $\pm$ must be the same in
(\ref{eseven}), (\ref{eeight}). This choice of signs stems from the
fact that in   $AdS_3\times S^3$ we
can reverse the sign of both angular coordinates $\phi, \psi$ while
leaving the $B^{RR}$ field unchanged, so this
sign change is a symmetry of the giant graviton DBI action on
$AdS_3\times S^3$. The radii $\bar\rho, \bar
\theta$ are {\it arbitrary}, so that the size of the giant graviton
in $AdS_3$ and in $S^3$  does not depend
on the angular momentum it carries; in fact the angular
momentum is determined from the charges $m_1, m_5$. (The fact that
the effective potential governing the radial size of
the giant graviton is flat in $AdS_3\times S^3$ was noted in the
early papers on giant gravitons \cite{giant}.)

Note that if we take the limit $\bar\rho\rightarrow 0$ then we get a
giant graviton expanding on $S^3$ only, and if we take
$\bar\theta\rightarrow \pi/2$ then we get a giant graviton expanding
on $AdS_3$ only.

\begin{figure}
\epsfysize=2.5in \epsffile{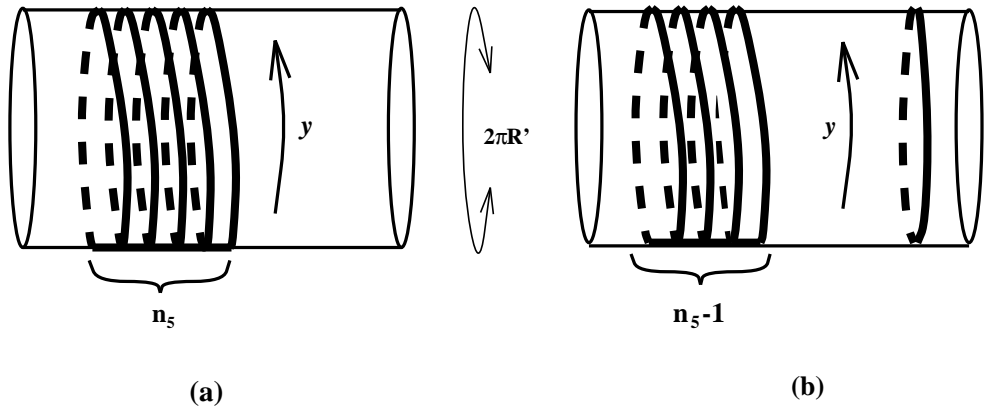}
\caption{\label{fig3}
(a) In all bound states the F string closes after $n_5$
turns around the $y$ direction. \quad (b)   One turn of the
F string is broken off; this gives a giant graviton.
}
\end{figure}

\subsection{The FP  representation of a  giant graviton}

In subsection 3.1 we had seen that the spacetime $AdS_3\times S^3$
was obtained under S,T dualities from a particular
configuration of the FP system -- the F string carried a traveling
wave in a form of a uniform helix executing a single turn.
Let us now see if we can find the FP representation of the spacetime
$AdS_3\times S^3$ which has present in it a single
giant graviton which we regard as a `test particle';  i.e., for the
moment we ignore its backreaction on the geometry.

Figure 3(a) depicts the $F$ string winding $n_5$ times around the $y$
circle before closing on itself; this is the
behavior of the F string for all bound states of the system.  In Figure
3(b) we have depicted {\it two} separate F strings.\footnote{Since we
have two F strings with different winding
numbers, we have drawn the strings in Fig 3(a,b) as multiwound strings
on $0<y<2\pi R'$ rather than the `opened up'
strings in Figs 1,2.} The first F string winds $n_5-1$ times around the
$y$ direction, and if we open it up to its full length, it
will have a  vibration profile similar to that of Fig 1(a) -- thus by itself
this string would generate the spacetime $AdS_3\times
S^3\times T^4$ after dualities. The second F string winds just once around the
$y$ circle and carries no vibrations; it is thus much lighter than
the first string and for the moment we regard it as a test
string in the background produced by the first F string.  The center
of the helix produced by the first  string is at $x_i=0,
i=1,2,3,4$. The second F string is placed at  some position
$x_i=\bar x_i$ in the transverse four dimensional space.

After the S,T dualities discussed in section 2  we get the D1-D5
geometry in the R sector.  The helical F string gives rise to
the $AdS_3\times S^3$ metric (\ref{esix}). The other F string is
being regarded as a test
string, so it does not contribute to the metric.  After the
dualities this F string becomes a D5 brane that wraps $T^4$, and thus
appears as a `string' in the remaining 5+1 dimensions.
This worldsheet of this `string' in these 5+1 dimensions can be parametrized as
\be\label{giantFlat}
x_i=\bar x_i, ~~y=R\sigma^1, ~~t=\sigma^0
\ee
After the change of coordinates (\ref{etwo}), (\ref{ethree}) the
embedding of this `string' is\footnote{In general the conditions
(\ref{giantFlat}) lead to
$\tilde\phi=\tilde {\bar\phi}$, $\tilde \psi=\tilde {\bar \psi}$,
but we can always shift angular
coordinates to make
$\tilde{\bar \phi}=\tilde {\bar \psi}=0$.} (we write $\chi=y/R$)
\be
\theta=\bar\theta, ~~r=\bar r,~~\chi=\sigma^1,  ~~t=\sigma^0,\quad
\tilde \phi=0,\quad \tilde \psi=0.
\ee
The string is still stationary, and so does not look yet look like a
giant graviton. But after the coordinate change (\ref{efour})
bringing us to the NS sector we get the metric (\ref{esix}) of
$AdS_3\times S^3$, and the `string' is described by
\be
\theta=\bar\theta, ~~r=\bar r, ~~\chi=\sigma^1, ~~\psi=-\sigma^1,
~~\phi=-\sigma^0, ~~t=\sigma^0
\ee
We see that we get  the giant graviton profile (\ref{eseven}),
(\ref{eeight}).\footnote{   Note that we have obtained the
configuration with the negative choice of sign in
(\ref{eseven}), (\ref{eeight}).
We cannot obtain the configuration with the other sign if we start
from a spacetime which goes over to flat
space at infinity, since the part of the metric which joined the
$AdS_3\times S^3$ region to flat infinity breaks
the symmetry which reverses the angles on $S^3$. But after we take
the limit (\ref{five}) we do have such a
symmetry, and so the conclusions that we reach with one choice of
sign are the same as for the configurations
with the other sign.}

\subsection{Geometry produced by a giant graviton.}

In the previous subsection we treated the giant graviton as a probe
string on the undeformed
$AdS_3\times S^3$ background. This is a traditional view of a giant
graviton, but the approximation
breaks down as the charge of the giant graviton becomes comparable
with the charge of the original
$AdS_3\times S^3$. In the FP representation this means that the
length of the ``short'' string
which became the giant graviton becomes comparable with
the length of the helical
string which produced $AdS_3\times S^3$. At this point the giant
graviton cannot be considered as a
test particle and one has to take into account its backreaction on
the geometry\footnote{The similar
problem for giant gravitons on $AdS_5\times S^5$ was addressed in
\cite{balGiant}.}.

We can have various distributions of giant gravitons and
each such distribution has a different metric. To illustrate how all such
metrics may be constructed we take a particular distribution of giant
gravitons and construct the corresponding metric.  Based on the discussion
of the above subsection, we start with the following profile of a F string
`broken' into two parts:
\bea\label{doubleProf}
&&{{\bf F}}(v)=a{\bf e}_1\cos\left(\frac{2\pi v}{\xi l}\right)+
a{\bf e}_2\sin\left(\frac{2\pi v}{\xi l}\right),\quad 0\le v<l\xi\\
&&{{\bf F}}(v)=b{\bf e}_1,\quad l\xi\le v<l
\eea
The first line of this equation gives a helical string which produces
$AdS_3\times S^3$, and
the second line gives rise to a bound state of several giant
gravitons made from the D5-brane; the location of
these giant gravitons is given as in  (\ref{giantFlat}).

As before the relations (\ref{CNMProf}) give the coefficient
functions in the metric.
After the change of coordinates (\ref{ethree}) we get  the following
functions for the chiral null model:
\bea\label{CNMIntone}
H^{-1}=1+\frac{Q\xi}{{r}^2+a^2\cos^2\theta}+\frac{Q(1-\xi)}{r^2+
a^2\sin^2\theta-
2b\sqrt{r^2+a^2}\sin\theta\cos{\tildr\phi}+b^2},\\
A_{\phi}=-\frac{a^2Q}{{r}^2+a^2\cos^2\theta}\frac{2\pi }{l}\sin^2\theta,\quad
K=\frac{Qa^2}{{r}^2+a^2\cos^2\theta}\frac{1}{\xi}\left(\frac{2\pi
}{l}\right)^2
\eea
Substituting these functions in the metric of D1--D5 system
(\ref{D1D5Chiral}) and going to the NS sector using (\ref{RtoNS})
we find the geometry which is produced by a giant graviton whose
location can be expressed parametrically in
terms of world sheet variables as\footnote{We have assumed $b<a$. For
$b>a$ we get $\sin\theta=1, r\ne 0$ and
we get a giant graviton expanding on the $AdS$ rather than on the sphere.}
\be
\theta=\arcsin\frac{b}{a}, ~~r=0, ~~\chi=\sigma^1, ~~\psi=-\sigma^1,
~~\phi=-\sigma^0, ~~t=R\sigma^0.
\ee
The resulting metric is quite complicated and we will not write it
down explicitly. Let us instead
do the same procedure we did in order to get the Aichelburg--Sexl
type solution (\ref{NewASmetr}):
there
instead of looking at a shock wave produced by a single particle on
the big diameter of the sphere
(whose angular coordinate was given by $\phi=\phi_0+t/R$) we
considered  a set of such particles
uniformly distributed over the diameter. This change implies that we perform an
    average over
    location $\phi_0$ in the harmonic functions.
Let us now follow a similar path  for the giant gravitons. Instead
of the profile
(\ref{doubleProf}) we consider a  profile where the second part of
the F string is further broken up into a large
number of identical segments, and the location of these segments is
distributed uniformly on a circle of
radius $b$ in the
$x_1-x_2$ plane. For convenience we just write this profile as
\bea
&&{{\bf F}}(v)=a{\bf e}_1\cos\left(\frac{2\pi v}{\xi l}\right)+
a{\bf e}_2\sin\left(\frac{2\pi v}{\xi l}\right),\quad 0\le v<l\xi\\
&&{{\bf F}}(v)=b(\cos\phi_0{\bf e}_1+\sin\phi_0{\bf e}_2),\quad l\xi\le v<l
\eea
where it is assumed that an average will be taken over $\phi_0$ in
the harmonic functions in (\ref{CNMProf}).
We then
get
\bea\label{HarmGGFl}
H^{-1}=1+\frac{Q\xi}{f_0}+\frac{Q(1-\xi)}{f_1},\quad
A_{\phi}=-\frac{a^2Q}{f_0}\frac{2\pi }{l}\sin^2\theta,\quad
K=\frac{Qa^2}{f_0}\frac{1}{\xi}\left(\frac{2\pi }{l}\right)^2
\eea
where
\be
f_0={r}^2+a^2\cos^2\theta,\quad
f_1=
\left[(r^2+a^2\sin^2\theta-b^2)^2+4r^2b^2\cos^2\theta\right]^{1/2}
\ee
    Let us note that unlike the Aichelburg--Sexl
type solution which had constant
dilaton in the near horizon limit, the solution (\ref{HarmGGFl}) has
a non-constant dilaton even in the
near horizon  region:
\be
e^{2\Phi}=\frac{a^2}{\xi}\left(\frac{2\pi
}{l}\right)^2\left[\xi+(1-\xi)\frac{f_0}{f_1}\right]^{-1}
\ee
Looking at the harmonic functions (\ref{HarmGGFl}) near infinity we
find the D1 and D5 charges of the solution:
\be
Q_1=\frac{Qa^2}{\xi}\left(\frac{2\pi }{l}\right)^2,\quad Q_5=Q
\ee
For later use it is convenient to rewrite the expression for $A_\phi$
from (\ref{HarmGGFl}) in terms of the charge
$Q_1$:
\be
A_{\phi}=-\frac{\xi Q_1}{f_0}\frac{l}{2\pi }\sin^2\theta=-\frac{\xi
Q_1}{f_0}\frac{n_5}{R }\sin^2\theta
\ee
where in the last step we used the relation between the length $l$ of
the F string appearing in the harmonic
functions and  the radius $R$ of the $y$ circle obtained after
dualizing to D1-D5:
\be
l=\frac{2\pi n_5}{R}
\ee
    It is convenient to introduce
the perturbation parameter
$q\equiv 1-\xi$ instead of $\xi$; thus $q=0$ gives the $AdS_3\times
S^3$ spacetime without any giant
gravitons. Then in the near horizon limit we obtain the following
solution\footnote{Here and below by $ds_E^2$
we denote the six dimensional metric  in the Einstein frame}:
\bea
ds_E^2&=&e^{\Phi}\left\{Q_5(1-q)\left[-\frac{r^2+a^2}{Q_1Q_5}dt^2+
\frac{r^2}{Q_1Q_5}dy^2+\frac{
dr^2}{r^2+a^2}\right]\right.
\nonumber\\
&+&Q_5(1-q)\left[d\theta^2+
\sin^2\theta(d{\tildr\phi}-\frac{dt}{R})^2+\cos^2\theta(d{\tildr\psi}-
\frac{dy}{R})^2\right]\\
&+&\frac{qf_0}{Q_1}(-dt^2+dy^2)+\frac{qQ_5f_0}{f_1}\left[d\theta^2+\frac{d
r^2}{r^2+a^2}\right]\nonumber\\
&+&\left.\frac{qQ_5}{f_1}\left[r^2\cos^2\theta
d{\tildr\psi}^2+(r^2+a^2)\sin^2\theta
d{\tildr\phi}^2\right]\right\}\\
B^{RR}_{t{\tildr\psi}}&=&-\frac{Q_5}{R}(1-q)\cos^2\theta,\quad
B^{RR}_{y{\tildr\phi}}=-\frac{Q_5}{R}(1-q)\sin^2\theta,\quad
B^{RR}_{ty}=\frac{f_0}{Q_1}\nonumber\\
B^{RR}_{{\tildr\phi}{\tildr\psi}}&=&\frac{Q_5(1-q)}{2}\cos
2\theta+\frac{Q_5q(r^2\cos
2\theta+b^2-a^2\sin^2\theta)}{2f_1}
\eea

The above solution was constructed in  the Ramond sector. We can now
perform the
spectral flow to the NS sector using the relation (\ref{RtoNS}).
Then we get the  metric
\bea
ds_E^2&=&e^\Phi\left\{Q_5(1-q)\left[-\frac{r^2+a^2}{Q_1Q_5}dt^2+\frac{
r^2dy^2}{Q_1Q_5}+\frac{
dr^2}{r^2+a^2}
+d\theta^2+
\sin^2\theta d{\phi}^2+\cos^2\theta d{\psi}^2\right]\right.\nonumber\\
&+&\frac{qf_0}{Q_1}(-dt^2+dy^2)+\frac{qQ_5f_0}{f_1}\left[d\theta^2+\frac{d
r^2}{r^2+a^2}\right]\nonumber\\
&+&\left.\frac{qQ_5}{f_1}\left[r^2\cos^2\theta
(d{\psi}+\frac{dy}{R})^2+(r^2+a^2)\sin^2\theta
(d{\phi}+\frac{dt}{R})^2\right]\right\}
\eea
which describes the metric created by the above discussed
distribution of giant gravitons in
$AdS_3\times S^3$.

In the above we have separated part of the F charge away from the FP
bound state to create the giant
gravitons, and this mapped after S,T dualities to giant gravitons
arising from D5 branes wrapped on the $T^4$.
We can also start by separating off momentum modes (P) from the FP
bound state, which implies that some of
the momentum exists as separate massless particles moving along the
$y$ direction instead of being the
momentum of traveling waves on the F string. The dualities map these
momentum modes to giant gravitons
created by D1 branes in $ADS_3\times S^3$.  We can also consider
giant gravitons that are bound states of D1
and D5 branes -- these arise from breaking the FP string into two
parts each of which have nonzero F and P
charges. For small F',P' charges in the separated part we can still
regard the corresponding giant graviton as
pointlike, but note that the momentum P' gives the F' string a finite
transverse size, which translates to a
nonzero transverse size for the D1-D5  giant graviton.  This is of
course the same phenomenon discussed in
\cite{lm4} which gives all D1-D5 bound states a nonzero size, and has
been the essential physics giving rise to
the  different metrics (\ref{D1D5Chiral}) for the D1-D5 bound state.

\section{Asymptotic behavior of the supergravity solution and its
relation to dispersions in the CFT dual}
\renewcommand{\theequation}{5.\arabic{equation}}
\setcounter{equation}{0}

The general solution (\ref{D1D5Chiral}) gives metrics for all R ground
states, or equivalently, all chiral primaries if we take the
limit (\ref{five}) and perform  spectral flow to the NS sector. In
this section we examine these metrics in the limit (\ref{five}), and
observe the leading corrections to the spacetime $AdS_3\times
S^3$. We will see that this analysis gives us a way to infer
certain characteristics of the CFT state from the asymptotic
behavior of the corresponding geometry.

The simplest geometries are the special class (\ref{two}); these
metrics are locally exactly $AdS_3\times S^3$. Thus at infinity
in $AdS$ space we find only a gauge field which is pure gauge,
and scalars like the dilaton $\Phi$ are fixed at their background
value. The corresponding CFT states (\ref{ChirPrCon}) are particularly
simple; all the twist operators $\sigma_n$ have the same order
and the same spin $(--)$. By contrast the Aichelburg-Sexl type
solution (\ref{NewASmetr}) is {\it not} locally $AdS_3\times S^3$, so if
we examine the metric near infinity we will find corrections that
fall off with various powers of $r$. The corresponding CFT state
(\ref{oneState}) has twist operators of different orders, so that we find
a {\it dispersion} in the values of the $n_i$.

We start with a more detailed analysis of the asymptotic behavior
of the Aichelburg-Sexl type solution (\ref{NewASmetr}), and then study the
asymptotic behavior of the general metrics (\ref{D1D5Chiral}) in the
limit (\ref{five}).
We observe
from this analysis that there is a close relation between the
existence of corrections at infinity and the presence of
dispersions in the orders and spins of the
$\sigma_n^{\epsilon\epsilon'}$ in the CFT state.

\subsection{Asymptotic behavior of The Aichelburg-Sexl type solution
(\ref{NewASmetr})}

The leading correction at $r\rightarrow\infty$ to the metric on
           $S^3$  is
\be
h_{ab}= \frac{qL^{2}}{r^{2}}\left(\begin{array}{ccc}0 & 0 & 0 \\ 0 & \cos^{4}
\theta & 0 \\ 0& 0 & -\sin^{4}\theta \end{array}\right)
\label{fone}
\ee
(We use indices $a,b\dots$ for $S^3$ coordinates and $\mu, \nu\dots$
for the $AdS_3$.)
We wish to put this perturbation into the basis which
diagonalizes the
quadratic part of the supergravity Lagrangian. Such a diagonalization
was performed for the
case $AdS_5\times S^5$ in \cite{van} and for the case $AdS_3\times S^3$ in
\cite{arut}. We write
\be
h=h_a^a, ~~h_{(ab)}=h_{ab}-{1\over 3}g_{ab}h, ~~\nabla^b h_{(ab)}=0
\ee
where all indices are raised and lowered with the background metric
$AdS_3\times S^3$,
and the last equation indicates  the choice of de-Donder gauge.
To bring the perturbation (\ref{fone}) to the de-Donder gauge we make
the diffeomorphism
\be
x^a\rightarrow x^a+\xi^a, ~~~\xi_\theta={1\over 4} {qL^2\over r^2}\sin 2\theta,
~~\xi_\phi=\xi_\psi=0
\label{ftwo}
\ee
We then get
\be
\tilde h_{ab}=h_{ab}+\nabla_a
\xi_b+\nabla_b\xi_a=(\frac{qL^2}{r^2}\cos 2\theta) g_{ab}
\ee
Thus the traceless part of the perturbation vanishes ($\tilde h_{(ab)}=0$), and
\be
\label{ffour}
\tilde h=\frac{3qL^2}{r^2}\cos 2\theta
\ee
so that $\tilde h$ is in the second harmonic on $S^3$.

Now we consider the perturbation of $B_{ab}^{RR}$. This perturbation is
\be
b=\frac{qL^{2}}{r^{2}}\sin^{2}\theta\cos^{2}\theta d\psi\wedge d\phi
\ee
The diffeomorphism (\ref{ftwo}) shifts this to
\be
\tilde{b}=\frac{2qL^{2}}{r^{2}}\sin^{2}\theta\cos^{2}\theta
d\psi\wedge d\phi\label{nnb}
\label{ffive}
\ee
We note that this field already satisfies the Lorentz gauge condition
$\nabla^a b_{ab}=0$
so we do not need to perform a further gauge transformation to bring
it to the form used in
\cite{arut}.

We now note that the perturbations $h, b$ mix with each other in the
quadratic supergravity action, and
the values (\ref{ffour}), (\ref{ffive}) are  precisely  those\footnote{The
normalization for the 2-form potential in \cite{arut} is such
that the action is $-{1\over 3} H^2$. We use instead the normalization that
is  conventional for the 10-dimensional IIB theory where the action is
$-{1\over 12} H^2$.} that give the scalar $\sigma$ in the
second harmonic. This scalar
          has \cite{arut}
\be
m_\sigma^2=k(k-2), ~~k \ge 1
\ee
Since we have $k=2$ we get $m^2=0$.
        From (\ref{ffour}) we see that the amplitude of this scalar falls off
at infinity as $1/r^2$, which is the correct falloff
         for  a massless scalar in $AdS_3$.

\subsection{Asymptotic behavior of the general metrics (\ref{D1D5Chiral})}

Now consider the general solution (\ref{D1D5Chiral}). We assume the limit
(\ref{five}) has been taken so that
for $r\rightarrow\infty$ we are looking near infinity of $AdS$ space
rather than infinity of flat space.

\subsubsection{ Dilaton and volume of $T^4$}

First we note that the scalars $\Phi$ (the dilaton) and $V$  (the
volume of $T_4$)
are both functions of
$H(1+K)$. Thus we expand $H(1+K)$ for large $r$ to see if
these scalars are excited.
We will use  Cartesian coordinates for a part of the following
analysis, but note  that at leading order
the radial coordinate of $AdS_3$ space is
\be
r^2=x_ix_i
\ee
and the angular variables describe  the $S^3$.

In the expressions (\ref{CNMProf}) we expand
\be
|x-F|^{-2}={1\over x^2}[1+ {2 x\cdot F\over x^2} - {F^2\over x^2}+{4
(x\cdot F)^2\over x^4}+\dots]
\label{fiveonetwo}
\ee
where the dots imply higher order terms in $1/r$. We define for any
function of $v$ the average
\be
{1\over l}\int_0^lS(v)dv\equiv \langle S\rangle
\ee
We then find
\be
H(1+K)=\langle(\dot F)^2\rangle+ {2 x_i\over x^2}[\langle(\dot
F)^2F_i\rangle- \langle F_i\rangle\langle(\dot
F)^2\rangle]+\dots
\label{fivenine}
\ee
       Note that if
\be
(\dot F)^2=constant
\label{fsix}
\ee
then the correction to $H(1+K)$ vanishes at  leading order.
We thus see that the above scalars are
excited at order $1/r$ at infinity if and only if we have a {\it dispersion} in
the quantity $(\dot F)^2$. Recall that the functions $F_i(v)$ give,
after Fourier transformation,
the orders and spins of the twist operators
$\sigma_n^{\epsilon\epsilon'}$ creating the dual
CFT state (eqn. (\ref{three})). Thus the dispersion of $(\dot F)^2$ is
directly related to
the dispersion
of the orders and spins of these twist operators. The metric
(\ref{two})  had the profile $F_i(v)$ in the form
of a uniform helix, and thus had $(\dot F)^2=constant$. The
Aichelburg-Sexl solution had a profile
$F_i(v)$ which was composed of helical as well as straight line
segments, which imply non--constant
$(\dot F)^2$. But the limits (ii') in section \ref{Sect33} imply an
`effectively constant' value for
$(\dot F)^2$ as far as its use in (\ref{fivenine}) is concerned, and
the dispersion in
(\ref{fivenine}) vanishes. This is in accord with the fact that the
above scalars are not excited in the
Aichelburg-Sexl solution.

\subsubsection{ The metric perturbation}

We now address the corrections  to the metric. Note that the
6-dimensional Einstein metric is obtained (upto an overall constant)
from the 10-dimensional
string metric
(\ref{D1D5Chiral}) by simply dropping the term $dz_\alpha dz_\alpha$;
this follows because
the 6-dimensional dilaton is constant for these solutions. We rewrite the
resulting metric in a form that gives the dimensional reduction to
$AdS_3$, since it is this form of the metric that will give the
excitations
dual to operators in the CFT. We get
\be
ds^2= \mu_{ij} (dx_i+C_i dt+D_i dy)(dx_j+C_j dt+D_j dy)+\nu_1
dt^2+\nu_2 dy^2+\sigma dt dy
\label{ften}
\ee
with
\bea
\mu_{ij}&=&\sqrt{1+K\over H}\delta_{ij}-\sqrt{H\over 1+K}
(A_iA_j-B_iB_j)\nonumber \\
C_i&=&\mu^{-1}_{ik} \sqrt{H\over 1+K} A_k, ~~D_i=\mu^{-1}_{ik}
\sqrt{H\over 1+K} B_k\nonumber \\
\nu_1&=&-\sqrt{H\over 1+K}-\mu_{ij} C_iC_j, ~~\nu_2=\sqrt{H\over
1+K}-\mu_{ij}D_iD_j\nonumber \\
\sigma&=&-{1\over 2} \mu_{ij} C_i D_j
\label{fel}
\eea
We will later convert the $x_i$ to $r$ and angular coordinates on
$S^3$; the coordinates $r,t,y$ will form the $AdS$ space.

\subsubsection{ The correction $\sigma dtdy$}

The special metrics (\ref{two}), (\ref{NewASmetr}) do not have any term mixing
$dt$ and $dy$, so we would like to understand more generally
what conditions imply a vanishing of $\sigma$ in the metric
(\ref{ften}).  Note that the vanishing of $dtdy$ mixing in the R
sector implies the vanishing of $dt d\chi$ mixing in the metric
obtained for  the NS
sector after dimensional reduction to  the $t,r,\chi$ spacetime.  To
leading order
$\mu_{ij}$ is proportional to
$\delta_{ij}$, so
\be
\sigma=0 ~\Rightarrow~ A_iB_i=0
\label{ftw}
\ee

Expanding near infinity we find
\be
A_i=-Q[{1\over x^2} \langle \dot F_i\rangle +\langle {2 x\cdot F\over x^4}\dot
F_i\rangle ]=-2Q\langle \dot F_iF_j\rangle {x_j\over x^4}
\label{uuone}
\ee
where we have used the fact that $F_i(v+L)=F_i(v)$, which implies that for
any quantity $S(v)$ made from the $F_i$
\be
\langle \dot S\rangle =0
\label{fthir}
\ee
We define
\be
a_{ij}=\langle \dot F_iF_j\rangle
\ee
Using (\ref{fthir}) we find
\be
a_{ij}=-a_{ji}
\ee
Thus we get at leading order
\be
A_i=-2Q a_{ij}{x_j\over x^4}
\ee
${}$From (\ref{DualFields}) we find
\be
B_i=-2Q b_{ij} {x_j\over x^4} , ~~b_{ij}={1\over 2} \epsilon_{ijkl} a_{kl}
\label{fsixt}
\ee
Since $a_{ij}$ is an antisymmetric matrix we can perform an $SO(4)$
rotation on the coordinates $x_i$ to bring it to the form
\be
a_{12}=-a_{21}=\alpha, ~~a_{34}=-a_{43}=\beta
\label{ffourt}
\ee
with all other components zero. The condition (\ref{ftw}) then
becomes, using (\ref{fsixt})
\be
a_{12}a_{34}=0  ~\Rightarrow~ a_{12}=0 ~\mbox{or} ~ a_{34}=0
\label{ffift}
\ee
If $a_{34}=0$ then we find that the profile $F_i(v)$ describes
vibrations only in the $x_1-x_2$ plane; thus the condition for
vanishing of
$\sigma$ is that there be no dispersion in the {\it orientation plane
} of the string in the FP description of the metric. Rotation in a
given plane corresponds by eqn. (\ref{theMap}) to the
spins $(\epsilon\epsilon')$ of all twist operators being in the same
wavefunction,  so in the CFT  state we will have
no dispersion
in these spins.\footnote{We have decided to do the spectral flow from
the R to the NS sector using a specific choice of
$U(1)\times U(1)$ out of the ${\cal R}$-symmetry group $SU(2)\times
SU(2)$; this choice was made in eqn. (\ref{etwo}),
(\ref{RtoNS}). The rotation that brings $a_{ij}$ to the form
(\ref{ffourt}) maps the set $x_1, x_2, x_3, x_4$ to a new
orthonormal set
$\hat x_1,
\hat x_2, \hat x_3, \hat x_4$. Thus in eqn. (\ref{ffift}) the
vanishing of $a_{34}$ is the vanishing of $a$ in the $\hat x_3\hat
x_4$ plane, though we did not indicate this in the equation to avoid
cumbersome notation. A string rotating in the
$x_1-x_2$ plane would imply spins $(-~-)$ for all twist operators, but
a string rotating along $\hat x_1-\hat x_2$ implies
that all twist operators have some wavefunction $(\epsilon,
\epsilon')$  obtained by rotating the spin $(-~-)$ by the
appropriate $SO(4)=SU(2)\times SU(2)$ group element.}

\subsubsection{ Corrections to the metric on $S^3$}

Consider the corrections to
\be
\mu_{ij}=\sqrt{K\over H}\delta_{ij}-\sqrt{H\over K} (A_iA_j-B_iB_j)
\label{fsevt}
\ee
We will consider separately the corrections to the two terms in
$\mu_{ij}$, and combine the effects at the end.
We find
\bea
\sqrt{1+K\over H} dx_idx_i={Q\over x^2}\langle (\dot F)^2\rangle
^{1/2}&+& {Q\over
x^2}\langle (\dot F)^2\rangle ^{-1/2}{x_i\over x^2}[\langle (\dot
F)^2 F_i\rangle + \langle (\dot F)^2\rangle
\langle F_i\rangle ]\nonumber\\
&{}&+{Q\over x^2}\langle (\dot F)^2\rangle ^{1/2}{x_ix_j\over x^4}T_{ij}+\dots
\label{feit}
\eea
with
\bea
T_{ij}&=&{1\over 2} \langle (1+{(\dot F)^2\over  \langle (\dot
F)^2\rangle })(- F^2  \delta_{ij}+
4 F_iF_j)\rangle \nonumber\\
&{}&~~-{1\over 2} [\langle F_i\rangle -\langle {(\dot F)^2\over \langle (\dot
F)^2\rangle }F_i\rangle ]~[\langle F_j\rangle -\langle {(\dot
F)^2\over \langle (\dot F)^2\rangle }F_j\rangle ]
\label{fnit}
\eea
The term of order $1/r^3$ in (\ref{feit}) can be canceled by a shift
\be
x_i=x'_i+v_i, ~~v_i={1\over 2} [\langle {(\dot F)^2\over \langle
(\dot F)^2\rangle } F_i\rangle +  \langle F_i\rangle ]
\ee
(We will henceforth drop the primes on the shifted coordinates to
avoid cumbersome notation.)
This shift adds to $T_{ij}$ a contribution
\be
\tilde T_{ij}=[v^2\delta_{ij}-4v_iv_j ]
\ee

We see that we get a simplification in the form of $T_{ij}$ if $(\dot
F)^2$ is constant; this is the same condition (\ref{fsix}) that led
to the
vanishing of dilaton and torus volume scalars.   In this case we get
\be
T^{total}_{ij}=T_{ij}+\tilde T_{ij}=4 [\langle F_iF_j\rangle -\langle
F_i\rangle \langle F_j\rangle ] -
\delta_{ij}[\langle F_kF_k\rangle -\langle F_k\rangle \langle F_k\rangle ]
\label{ftwenty}
\ee
We see that the above expression is also written as a sum of
dispersions, but these dispersions cannot all vanish in any
configuration since that would imply that  the $F_i$ be constants.

\subsection{Asymptotic fields for constant $(\dot F)^2$}

We now write down expressions for the asymptotic form of some metric
components, for the case where there is no dispersion in  $(\dot F)^2$.

\subsubsection{The metric on $S^3$}

We have analyzed above the first contribution to $\mu_{ij}$ in
(\ref{fsevt}), and we now compute the contribution
of the gauge fields $A_i, B_i$.
       We assume that the matrix $a_{ij}$ has been put in the form
(\ref{ffourt}).
We map the $x_i$ to polar coordinates using (\ref{EpolarMap}).
We find
\bea
ds^2_{gauge}&\equiv&-\sqrt{H\over K} (A_iA_j-B_iB_j) dx_idx_i\nonumber \\
&=&{Q\over |{\dot F}|r^2}\left[-(\alpha \sin^2\theta d\phi+\beta
\cos^2\theta
d\psi)^2+(\beta \sin^2\theta d\phi-\alpha\cos^2\theta d\psi)^2
\right]
\label{ftwone}
\eea
  From (\ref{uuone}), (\ref{fthir}) we see that the nonvanishing of the
$A_i, B_i$ at infinity can be itself regarded
as arising from a nonzero value for the dispersion
\be
\langle \dot F_i F_j\rangle - \langle \dot F_i\rangle \langle
F_j\rangle = \langle \dot F_i F_j\rangle
\ee

To put the metric in de-Donder gauge we make a diffeomorphism similar
to (\ref{ftwo})
\be
\xi_\theta=Q{\alpha^2-\beta^2\over 4|{\dot F}|r^2}\sin2\theta,
\quad \xi_\phi=\xi_\psi=0
\ee
which yields
$$
ds_{gauge}^2=Q{\alpha^2-\beta^2\over |{\dot F}|r^2}
\left[\cos 2\theta (d\theta^2+\cos^2\theta d\psi^2+\sin^2\theta d\phi^2)-
{4\alpha\beta\over \alpha^2-\beta^2}\sin^2\theta\cos^2\theta
d\phi d\psi\right]
$$

Combining the contributions from (\ref{feit}) and $ds^2_{gauge}$ we
get for the  metric on $S^3$
\bea
ds^2_{sphere}&=&Q|\dot F| d\Omega^2+\left[{Q|\dot F|\over
r^2}\left[4n_in_j \langle F_iF_j\rangle -\langle F^2\rangle
\right]+{Q(\alpha^2-\beta^2)\over |{\dot F}|r^2}\cos
2\theta\right]d\Omega^2\nonumber\\
&-&{4Q\alpha\beta \over |{\dot F}|r^2}\sin^2\theta\cos^2\theta d\phi d\psi
\eea
where $n_i={x_i\over x}$.

We find that just as in the case of the Aichelburg--Sexl metric we
get a contribution to trace $h$   in the
second harmonic, which is dual to an operator with $\Delta=2$ in the
CFT.\footnote{In  general we can also find
a contribution to $h$ in the zeroth harmonic on $S^3$, with this
contribution falling like
$1/r^2$ at infinity.  This power law suggests a massless scalar, but
the quadratic supergravity action does not have such a scalar made
from $h$. But the
theory has cubic couplings, for example
$h\partial\phi\partial\phi$ where $\phi$ is the dilaton. We can have
$\phi\sim 1/r$, and also $\partial \phi\sim
1/r$ where the derivative is taken along the $S^3$. We then get from
the cubic coupling a contribution $h\sim
1/r^2$, with $h$ in the zeroth harmonic. Thus we have to be careful
about higher order terms in
the Lagrangian in the analysis of asymptotic fields.}

\subsubsection{Gauge fields on $AdS_3$ }

Now we look at the asymptotic behavior of terms in the metric that
mix the $S^3$ coordinates with the $AdS_3$ coordinates. This part
of the metric turns out to be
\be
ds_3^{mixed}={2\over |{\dot F}|}\left[
(\alpha \sin^2\theta d\phi+\beta \cos^2\theta d\psi)dt+(\beta
\sin^2\theta d\phi-\alpha\cos^2\theta d\psi)d\chi\right]
\ee
We note that since $h^\phi_\mu,h^\psi_\mu$ do not depend upon sphere
coordinates and $h^\theta_\mu=0$
(here $\mu$ is an AdS index), the metric $ds_3^2$ is automatically in the gauge
\be
\nabla_a h^a_\mu=\d_a h^a_\mu-\Gamma_{ab}^bh^b_\mu=0
\ee
We further find
\be
\nabla^2 h_{a\mu}=-2h_{a\mu},
\ee
(where the Laplacian is taken over the sphere).  Comparing with the fields
studied in
\cite{arut} we see that such an excitation is in the  harmonic $k=1$.
The field $h_{a\mu}$ mixes with $B^{RR}_{a\mu}$, and
the lightest field arising from this combination  has scaling
dimension $\Delta=1$ in the CFT.

\section{Discussion}
\renewcommand{\theequation}{6.\arabic{equation}}
\setcounter{equation}{0}

Let us summarize our main conclusions.
In most treatments of $AdS/CFT$ duality we start with the geometry
$AdS_m\times S^n$ which is dual to the vacuum of the CFT.
We then consider excitations over this vacuum perturbatively, for
example in the computation of multipoint  correlation
functions of chiral primaries \cite{witten, freedman}. For the case
$AdS_3\times S^3$ however we are able to write down exact
supergravity
solutions that describe arbitrary chiral primary states. These
geometries are {\it not} small perturbations to  $AdS_3\times S^3$,
though they go to $AdS_3\times S^3$ at large $r$.  The availability
of these solutions allows us to probe questions that we cannot address
at present using other $AdS_m\times S^n$ spaces.

At first it appears to be  clear what a chiral primary is: we have
one or more massless quanta of supergravity
rotating around a diameter of $S^3$.   But a further look indicates
that the problem is more complex: we can find
more than one family of metrics that have  $\Delta=J$. Further, the
work on giant gravitons suggests that
the rotating quanta may themselves be finite size objects rather than
pointlike particles, and this would create more
complicated metrics after the backreaction is taken into account.

We found that two special families of metrics with $\Delta=J$ -- the
conical defect metrics and the Aichelburg-Sexl type solution --
were special cases of the general set of chiral primaries that can be
written down using the map to the FP system developed in \cite{lm4}.
But the CFT states dual to these two families were characterized by
rather different distributions of twist operators -- in the first
case all
twist operators were identical while in the second case there was a
wide dispersion in their orders $n_i$.

The giant graviton solutions turned
out
to not describe states of the bound D1-D5 system at all -- the bound
states map under S,T dualities to a single F string
carrying vibrations, while the giant gravitons correspond to breaking
this string into two or more pieces. It is noteworthy that a giant
graviton type solution can be written down not only for the space
$AdS_3\times S^3$ but also for all the geometries (\ref{D1D5Chiral}) which go
over at
infinity to flat space. This is easy to see: in the R sector we have
to break off some of the D1 and D5 branes  and separate them from the
rest of the D1-D5
bound state by a transverse displacement.  This solution always
exists classically since for vanishing moduli there is no force
between the D1-D5
bound state and  transversely displaced D1 or D5 branes. The
coordinate change (\ref{RtoNS}) to the NS sector then
makes this static D1 brane rotate, and we get a D1 brane that maintains its
finite size while satisfying its classical equation of motion.

It is important to note that such a giant graviton, while appearing
to be a supersymmetric solution at classical order,  is not a
supersymmetric solution
in the exact theory.  To see this consider for  simplicity two
parallel D-p branes, each wrapped on a finite torus $T^p$  of volume
$V$ and placed
a distance
$a$ apart. Classically this configuration has a mass $M\sim {V\over
g}$ where $g$ is the string coupling, and a charge that is related to
the
mass by a BPS condition. But in order for the two branes to be
localized at their respective positions we must make a superposition
of their
momentum eigenfunctions, and this adds in a `localization energy' of
order ${p^2\over 2M}\sim {g\over a^2V}$.  Thus to get mass equaling
charge we must either place the two branes in their zero momentum
eigenfunctions -- in which case their mean separation becomes
infinity and we do not get a set of branes at finite separation $a$
-- or we take their (unique) bound state, where the wavefunction
describing their relative separation is a very particular wavefunction
and the exact solution is known to be BPS.

The above discussion extends to the geometry that we can create by
putting $N$  branes at each of the locations $x=\pm {a\over 2} $.
Classically this geometry will appear supersymmetric, but for any
finite $N$ it is clear that corrections of order $g$ will have to
show us a
breakdown of supersymmetry. (In a worldsheet description this
breakdown should occur due to one loop open string diagrams.) Similarly we can
conclude that the giant graviton solutions do not represent exactly
supersymmetric states and are thus not chiral primaries.\footnote{We
have taken the F string in the FP solution to be described by a
classical vibration profile. This is a coherent state excitation and
thus not an
energy eigenstate, but in principle we can consider the energy
eigenstates instead and then we would have exact chiral primary
states. If
we try to make energy eigenstates out of the giant graviton solutions
then the giant gravitons delocalize all the way to infinity in $AdS_3$
since the potential is flat;  such a delocalization does not happen
for the bound states which are described by a single F string carrying
momentum P.} We note that the case of $AdS_3\times S^3$ is rather
special however, and this conclusion may not carry over to giant
gravitons in
other $AdS_m\times S^n$ spaces.

Returning to the D1-D5 bound state solutions, we recall that the
prescription of \cite{witten} allowed us to compute expectation
values of
operators in the NS vacuum of a CFT by using the dual supergravity.
It appears plausible that given the solutions representing chiral
primary states, there would be a way to compute expectation values of
operators in such states, by a perturbative  supergravity
computation around these more complicated backgrounds.  Even without
setting up such a computational scheme, it appears reasonable
that the behavior of the supergravity solutions near infinity
contains information about the corresponding CFT state:  for any
supergravity
mode the solution growing at $r\rightarrow\infty$ describes operators
inserted at the boundary while the solution decaying at infinity
describes  the `state' of the CFT \cite{bala}.

In view of these expectations we have examined the
$r\rightarrow\infty$ behavior of the general supergravity solution
describing a
chiral primary.  We noted that the conical defect solutions were
locally $AdS_3\times S^3\times T^4$, so there were no excitations at
infinity
apart from a pure gauge term mixing the $S^3$ with the $AdS_3$. The
corresponding CFT state was created by a set of twist operators that
were all identical. The Aichelburg-Sexl type solution did have
nontrivial excitations of field at infinity, and there was a
dispersion in the
orders of the twist operators. We found that this was somewhat of a
general pattern -- we could relate the strength of several fields at
infinity to
some {\it dispersion} in the twist operators creating the state. It
would be interesting to see if such a principle holds more generally
in the
AdS/CFT correspondence.

\section*{Acknowledgements}

We would like to thank Sumit Das, Juan Maldacena and Inyong Park for
helpful discussions.
The work of SDM and AS was  supported in part by DOE grant
DE-FG02-91ER-40690.  OL was supported by NSF grant
PHY--0070928.
\appendix
\section{A chiral null model approach to D1-D5 solutions}
\label{AppGener}
\renewcommand{\theequation}{A.\arabic{equation}}
\setcounter{equation}{0}

In this appendix we will recall the construction of \cite{lm5} which
related solutions for the
vibrating string with solutions describing the D1--D5 system.

We begin with a fundamental string which carries momentum. The general
solution for such a string is
given by a chiral null model \cite{chnull}:
\bea\label{ChiralNull}
ds^2&=&H'({\vec x'},v')\left(-du'~dv'+K'({\vec x'},v')d{v'}^2+
2A'_i({\vec x'},v')dx'_i dv'\right)+
d{\vec x'}\cdot d{\vec x'}+d{\vec z'}d{\vec z'},\nonumber\\
B_{uv}&=&-G_{uv}=\frac{1}{2}H'({\vec x'},v'),\qquad
B_{vi}=-G_{vi}=-H'({\vec x'},v')A'_i({\vec x'},v'),\\
&&\qquad\qquad e^{-2\Phi}=H'^{-1}({\vec x'},v').\nonumber
\label{ChiralSolution}
\eea
Regarding $A'_i$ as a gauge field we can construct the field strength
${\cal F}_{ij}=A'_{j,i}-A'_{i,j}$. The
functions in the chiral null model are required satisfy the equations
\be\label{NullEqn}
\partial^2 H'^{-1}=0,\qquad \partial^2 K'=0,\qquad \partial_i{\cal F}^{ij}=0.
\ee
Here
$\partial^2$ is  the Laplacian
in the $x'_i$ coordinates. Note that the indices $i,j$ span the
subspace $\{ x_i\}$ where the metric is just
$\delta_{ij}$, and thus these indices are raised and lowered by this
flat metric.

We can now perform the chain of dualities which relates the vibrating
string with D1--D5 system:
\be\label{dualMap}
\begin{array}{|c|c|c|c|c|c|c|c|c|c|c|}
\mbox{P}&&\mbox{P}&&\mbox{P}&&\mbox{P}&&\mbox{F1}&&\mbox{D1}\\
&\stackrel{\textstyle S}{\rightarrow}&&
\stackrel{\textstyle T6789}{\longrightarrow}&&
\stackrel{\textstyle S}{\rightarrow}&&
\stackrel{\textstyle T56}{\longrightarrow}&&
\stackrel{\textstyle S}{\longrightarrow}&\\
\mbox{F1}&&\mbox{D1}&&\mbox{D5}&&
\mbox{NS5}&&\mbox{NS5}&&\mbox{D5}
\end{array}
\ee
Note that before performing the S duality relating (P,D5) and (P,NS5)
we also perform an
electric--magnetic duality (see \cite{lm5} for details). The
resulting D1--D5 geometry is:
\bea\label{AD1D5Chiral}
ds^2&=&\sqrt{\frac{H}{1+K}}\left[-(dt-A_idx^i)^2+(dy+B_idx^i)^2\right]
+
\sqrt{\frac{1+K}{H}}d{\vec x}\cdot d{\vec x}\nonumber\\
&+&\sqrt{H(1+K)}d{\vec z}\cdot d{\vec z}\\
e^{2\Phi}&=&H(1+K),\qquad
C^{(2)}_{ti}=\frac{B_i}{1+K},\qquad
C^{(2)}_{ty}=-\frac{K}{1+K},\nonumber\\
C^{(2)}_{iy}&=&-\frac{A_i}{1+K},\qquad
C^{(2)}_{ij}=C_{ij}+\frac{A_iB_j-A_jB_i}{1+K}
\eea
The functions $H$, $K$ and $A_i$ appearing in this solution have the same
values as $H'$, $K'$ and $A'_i$:
\be
H({\vec x})=H'({\vec x'}), \qquad K({\vec x})=K'({\vec x'}),\qquad
A_i({\vec x})=A_i'({\vec x'}),
\ee
and the forms $B_i$ and $C_{ij}$ are defined by
\be\label{ADualFields}
dC=-^*dH^{-1},\qquad dB=-^*dA.
\ee
Here Hodge dual is taken with respect to the four dimensional space
$x^1,x^2,x^3,x^4$ with flat metric. Note that due to the equations of motion
for the null chiral model (\ref{NullEqn}):
\be
d^*dH^{-1}=0, \qquad d^*dA=0
\ee
the equations (\ref{DualFields}) can be integrated to give the forms $C$ and
$B$.

If we consider the metric created by a  single elementary string, then the
parameters of the chiral null model
are not arbitrary, but they are determined in terms of the vibration
profile ${\bf
F}(v)$:
\bea
H^{-1}=1+\frac{Q}{l}\int_0^l\frac{dv}{({\bf x}-{{\bf F}})^2},\quad
K=\frac{Q}{l}\int_0^l\frac{|\dot F|^2dv}{({\bf x}-{{\bf F}})^2},\quad
A_i=-\frac{Q}{l}\int_0^l\frac{{\dot F}_i dv}{({\bf x}-{{\bf F}})^2}
\label{qqqqone}
\eea
As an example we consider the helical profile:
\be
F_1=a\cos\frac{2\pi mv}{l},\quad F_2=a\sin\frac{2\pi mv}{l},\quad
F_3=F_4=0.
\ee
For this profile it is convenient to make a change of coordinates
(\ref{etwo}), (\ref{ethree}). Then
in terms of
the new coordinates $r,\theta,{\tildr\phi},{\tildr\psi}$ we find the
geometry of the
D1--D5 system \cite{mm}:
\bea\label{D1D5Ap}
ds^2&=&-\frac{{f_0}}{\sqrt{f_1f_5}}(dt^2-d{
y}^2)+ \sqrt{f_1f_5}(\frac{dr^2}{r^2+{ a}^2}+d\theta^2)+
\sqrt{\frac{{f_1}}{{f_5}}}\sum_{i=1}^4 dz^i dz^i\nonumber\\
&+&\frac{\sqrt{f_1f_5}}{f_0}\left[(r^2+\frac{{ a}^2Q_1Q_5\cos^2\theta}
{f_1f_5})\cos^2\theta d{\tildr\psi}^2
+(r^2+{ a}^2-\frac{{ a}^2Q_1Q_5\sin^2\theta}{f_1f_5})
\sin^2\theta d{\tildr\phi}^2\right]\nonumber\\
&-&\frac{2Q_1Q_5{a}}{\sqrt{f_1f_5}}\left[\sin^2\theta dtd{\tildr\phi}+
\cos^2\theta d{ y}d{\tildr\psi}\right]\\
e^{2{\tildr\phi}}&=&\frac{{ f}_1}{{ f}_{5}},\qquad
C^{(2)}_{ty}=-\frac{2Q_1}{{ f}_{1}}, \qquad
C^{(2)}_{t{\tildr\psi}}=-\frac{\sqrt{Q_1Q_5}{ a}\cos^2\theta} {{ f}_{1}},
\nonumber\\
\label{FNSend}
C^{(2)}_{y{\tildr\phi}}&=&-\frac{\sqrt{Q_1Q_5}{ a}
\sin^2\theta}{{ f}_{1}},
\qquad
C^{(2)}_{{\tildr\phi}{\tildr\psi}}=Q_5\cos^2\theta+
\frac{Q_5{ a}^2\sin^2\theta\cos^2\theta}
{{ f}_{1}}.
\eea
Here we have introduced three convenient functions:
\be\label{f0Def}
{ f}_0=r^2+{ a}^2\cos^2\theta,\qquad f_1=f_0+Q_1,\qquad f_5=f_0+Q_5.
\ee
and the charges are
\be
Q_5=Q,\quad Q_1=Q\left(\frac{2\pi am}{l}\right)^2
\ee
In the near horizon limit ($r\ll (Q_1Q_5)^{1/4}$) the metric reduces to
(\ref{two}) with
$\gamma=\frac{1}{m}$.

\end{document}